\documentclass[twocolumn]{article}
\usepackage[T1]{fontenc}
\usepackage[utf8]{inputenc}
\usepackage{amsmath, amssymb, physics, bm}
\usepackage{graphicx, xcolor}
\usepackage{booktabs}
\usepackage{caption}
\usepackage{amsthm}
\newtheorem{theorem}{Theorem}

\usepackage[margin=1.8cm, columnsep=0.6cm]{geometry}
\usepackage{authblk} % <-- Added for affiliations

\usepackage[numbers,sort&compress]{natbib}
\usepackage[colorlinks=true, linkcolor=blue, 
            citecolor=blue, urlcolor=blue]{hyperref}


\begin{document}

%--------------------------------------------------------------
\title{\textbf{Universal Description of Decoherence in Scale-Invariant Environments}}

\author[1]{Carlos Argüelles}
\author[2]{Gabriela Barenboim}
\author[1,3]{Gonzalo Herrera}
\author[1]{Tanvi Krishnan}
\author[2]{Héctor Sanchis}

\affil[1]{Harvard University, Department of Physics and Laboratory for Particle Physics and Cosmology, Cambridge, MA 02138, USA}
\affil[2]{Departament de F\'isica Te\'orica and IFIC, Universitat de Val\`encia-CSIC, E-46100, Burjassot, Spain}
\affil[3]{Kavli Institute for Astrophysics and Space Research, Massachusetts Institute of Technology, Cambridge, MA 02139, USA}

\maketitle
%--------------------------------------------------------------

\begin{abstract}
When a quantum system couples to a scale-invariant environment,
what form must its decoherence take?
We prove that the answer is unique: under locality, Lorentz
invariance, unitarity, and continuous scale invariance, the effect of any such environment is mathematically equivalent to that of an \emph{unparticle bath}~\cite{Georgi:2007ek} --- a scale-invariant continuum of states --- characterized entirely by the scaling dimension $d_{\mathcal{U}}$ of the coupled operator.
This is not a modelling choice but a consequence of conformal
symmetry.
All decoherence and dissipation exponents are fixed by
$d_{\mathcal{U}}$ through exact consistency relations, providing
falsifiable predictions independent of microscopic details.
We validate the framework using multi-channel transport data
from the unitary Fermi gas, where two genuinely independent observables yield a consistent $d_{\mathcal{U}} = 7/4$.
We further show that quantum Ising criticality, inflationary
cosmology, and high-energy astrophysical neutrinos---spanning
more than 25 orders of magnitude in energy---are unified
as specific realizations of the same structure.
A decoherence phase transition at $d_{\mathcal{U}} = 5/2$,
where quantum coherence is \emph{protected} rather than
destroyed at long times, is a qualitative prediction
inaccessible to any memoryless dynamical description.
\end{abstract}

%--------------------------------------------------------------
\section*{Introduction}
%--------------------------------------------------------------

Decoherence---the irreversible loss of quantum coherence through
environmental entanglement---underpins the quantum-to-classical
transition and sets fundamental limits on quantum technologies.
A central open question is: \emph{what constraints does symmetry
place on the form of decoherence?}

For generic environments the answer is: very few.
The Lindblad formalism~\cite{Lindblad1976} ---the standard framework for open quantum dynamics--- treats dissipation rates as free parameters,
while the Caldeira-Leggett model~\cite{CaldeiraLeggett1983} ---which models the environment as a bath of harmonic oscillators---
engineers spectral densities to match desired phenomenology.
These approaches capture enormous variety but obscure universal
structure.

Here we identify one case in which symmetry is maximally
constraining: when the environment is \emph{scale-invariant}.
Scale-invariant environments arise ubiquitously---at quantum
critical points in condensed matter, in the approximately de Sitter
geometry of inflationary cosmology, and potentially in quantum
gravity.
Despite this prevalence, the consequences for open-system dynamics
have not been systematically established.

We prove that under physically natural assumptions, a
scale-invariant environment coupled to any quantum system
\emph{must} be described by an unparticle bath~\cite{Georgi:2007ek}.
Originally proposed as speculative beyond-Standard-Model
phenomenology, unparticles are here elevated to a
\emph{universal characterization} of quantum systems coupled to scale-invariant environments.
This universality has a sharp experimental consequence: 
the power-law exponents of multiple independently measured observables  are not free parameters
but are fixed by a single number $d_{\mathcal{U}}$, obeying exact consistency relations that can be tested and falsified.

%--------------------------------------------------------------
\section*{The Uniqueness Theorem}
%--------------------------------------------------------------
What can we actually know about an environment we cannot directly observe,
armed only with the knowledge that it has no intrinsic scale?
Surprisingly, the answer is: almost everything.
The argument below is a uniqueness proof---we show not merely that
unparticle baths \emph{can} describe scale-invariant environments,
but that they \emph{must}.
The chain of reasoning is tight: scale invariance forces conformal
symmetry, conformal symmetry fixes the correlators of environmental fluctuations, and fixed
correlators determine frequency dependence of environmental noise completely.
No free functions, no adjustable spectral shapes remain.
A single number, the scaling dimension $d_{\mathcal{U}}$, determines
all decoherence and dissipation exponents exactly.
The proof is short but technical; the experimental consequences are developed in the following sections independently of its details.

\begin{theorem}[Unparticle Universality]
\label{thm:main}
Let a quantum system $S$ couple locally to an environment $E$
in $d$ spatial dimensions.
Assume: (i) $E$ exhibits exact continuous scale invariance;
(ii) the theory is Lorentz-invariant; (iii) the coupling is
local, $H_{\mathrm{int}} = g\,A_S(x)\,\mathcal{O}_E(x)$;
(iv) the full system evolves unitarily.
Then:
\begin{enumerate}
    \item $E$ is described by a conformal field theory (CFT).
    \item The noise spectrum of environmental fluctuations, $J(\omega)$, takes the unique form
    \begin{equation}
              J(\omega) = A\,\omega^{2\Delta - d - 1},
              \label{eq:spectral}
    \end{equation}
    where $\Delta$ is the scaling dimension of
    $\mathcal{O}_E$\footnote{Throughout this work we adopt the convention
\begin{equation}
    J(\omega) = -2\,\mathrm{Im}\,G_R(\omega,\mathbf{k}=0),
    \quad \omega > 0,
    \label{eq:J_convention}
\end{equation}
where $G_R(\omega)$ is the retarded Green's function of the
environmental operator $\mathcal{O}_E$, evaluated at zero
spatial momentum appropriate for a local coupling.
The overall normalization constant $A$ in Eq.~\eqref{eq:spectral}
absorbs the coupling strength $g^2$ and the CFT coefficient
$C_{\mathcal{O}}$; its precise value depends on the
microscopic theory and is not fixed by scale invariance alone.
Since all physical predictions of the framework depend only
on the \emph{scaling exponent} $2d_{\mathcal{U}}-3$ and not
on $A$, we do not specify a normalization convention beyond
Eq.~\eqref{eq:J_convention}, keeping the analysis general
and independent of microscopic details.}.
    \item This is equivalent to an unparticle bath with
          \begin{equation}
              d_{\mathcal{U}} = \Delta - \tfrac{d-2}{2}.
              \label{eq:dU}
          \end{equation}
    \item All dynamical exponents are uniquely determined
          by $d_{\mathcal{U}}$ (Table~\ref{tab:exponents}).
\end{enumerate}
\end{theorem}

The proof follows in four steps.
\emph{Step 1.}
In a relativistic QFT, continuous scale invariance combined with
Lorentz invariance and energy-momentum conservation implies full
conformal invariance~\cite{Polchinski1988}\footnote{In $d=2$, this follows rigorously~\cite{Polchinski1988}; in $d \geq 3$ it holds under the additional assumption of unitarity and absence of a virial current, supported by strong evidence in $d=4$~\cite{Nakayama2015,Luty2013}.}: tracelessness of
$T^{\mu\nu}$ and Poincar\'{e} symmetry together generate the full
conformal group $SO(d+1,1)$.
\emph{Step 2.}
Conformal Ward identities fix the two-point function of any
primary operator $\mathcal{O}_E$ with dimension $\Delta$
up to normalization:
$\langle \mathcal{O}_E(x)\mathcal{O}_E(0)\rangle =
C_{\mathcal{O}}/(x^2_E)^{\Delta}$.
There is no freedom to choose an alternative functional form.
\emph{Step 3.}
Analytic continuation to Lorentzian signature and Fourier
transformation at zero spatial momentum give the retarded
Green's function $G_R(\omega) \propto (-i\omega)^{2\Delta-d}$,
yielding $J(\omega) = -2\,\mathrm{Im}[G_R(\omega)]
\propto \omega^{2\Delta - d - 1}$.
\emph{Step 4.}
Matching to the unparticle spectral form
$\rho_{\mathcal{U}}(\omega) \propto \omega^{2d_{\mathcal{U}}-3}$
gives Eq.~\eqref{eq:dU}, and the memory kernels follow by
Fourier transformation (Table~\ref{tab:exponents}).

The theorem is not merely that unparticles provide a
\emph{convenient} parametrization: they are the \emph{only}
possible description.

\begin{table}[t]
\centering
\caption{%
\textbf{Complete set of scaling exponents for any
scale-invariant environment.}
All quantities are determined by the single parameter
$d_{\mathcal{U}}$, providing consistency relations that can be
independently measured and tested.
}
\label{tab:exponents}
\begin{tabular}{lcc}
\toprule
\textbf{Observable} & \textbf{Scaling} & \textbf{Exponent} \\
\midrule
Spectral density & $J(\omega)\propto\omega^{s}$
    & $s = 2d_{\mathcal{U}}-3$ \\
Dissipation kernel & $\eta(t)\propto t^{-\alpha_\eta}$
    & $\alpha_\eta = 2d_{\mathcal{U}}-2$ \\
Noise kernel (high-$T$) & $\nu(t)\propto T\,t^{-\alpha_\nu}$
    & $\alpha_\nu = 2d_{\mathcal{U}}-3$ \\
Damping function & $\Gamma_{\mathrm{damp}}\propto t^{\beta}$
    & $\beta = 3-2d_{\mathcal{U}}$ \\
Decoherence functional & $\Gamma_{\mathrm{decoh}}\propto t^{\gamma}$
    & $\gamma = 5-2d_{\mathcal{U}}$ \\
\bottomrule
\end{tabular}
\end{table}

\paragraph{Falsifiability and consistency relations.}

The exponents in Table~\ref{tab:exponents} are not independent.
They satisfy exact algebraic relations:
\begin{align}
    s + \gamma_{\mathrm{decoh}} &= 2, \label{eq:rel1}\\
    \alpha_\eta + \delta_{\mathrm{decoh}} &= 2, \label{eq:rel2}\\
    \alpha_\nu + \beta_{\mathrm{damp}} &= 0, \label{eq:rel3}
\end{align}
where $\delta_{\mathrm{decoh}} = 4 - 2d_{\mathcal{U}}$ is the
instantaneous decoherence rate exponent.
These are predictions, not fits: each relation connects an independently measurable bath property to an independently measurable system response, so neither side is derived from the other.
Measuring any two exponents independently extracts $d_{\mathcal{U}}$
from each; consistency tests the scale-invariance assumption.
\emph{Inconsistency falsifies scale invariance} and reveals
intrinsic scales, non-locality, or multiple competing sectors
in the environment.

A genuine CFT bath is further distinguished from a phenomenological
power-law mimic by the absence of a UV cutoff function modifying
the spectral density.
A generic oscillator bath with $J(\omega) \propto \omega^s
f(\omega/\Lambda)$ reproduces the same exponent but introduces
a rolloff at scale $\Lambda$; the unparticle bath has
$f \equiv 1$ over the entire scaling window.
This can be tested by checking that the consistency
relations~\eqref{eq:rel1}--\eqref{eq:rel3} hold simultaneously
across the full frequency range of the experiment.

\paragraph{Loopholes.}
The theorem requires \emph{continuous} scale invariance.
Physical deviations arise from: (i) infrared cutoffs (finite
temperature $T$, system size $L$, mass gap $m$) breaking scale
invariance below $\omega_{\mathrm{IR}}$; (ii) ultraviolet cutoffs
(e.g., lattice spacing, Planck length) above $\omega_{\mathrm{UV}}$;
(iii) discrete scale invariance (e.g., Efimov states in ultra cold atoms, hierarchical
models), which produces log-periodic modulations;
(iv) multiple unparticle sectors summing with different
$d_{\mathcal{U}}^{(i)}$, causing crossover behavior.
The unparticle description is valid within the scaling window
$\omega_{\mathrm{IR}} < \omega < \omega_{\mathrm{UV}}$.
Crucially, thermal corrections modify amplitudes but not
exponents: both the vacuum ($t \ll \beta$) and thermal
($t \gg \beta$) regimes exhibit power-law behavior with the
\emph{same} $d_{\mathcal{U}}$, only the prefactor changes.

%--------------------------------------------------------------
\section*{Experimental Validation:\\ Unitary Fermi Gas}
%--------------------------------------------------------------
The unitary Fermi gas ---a strongly interacting gas of cold fermionic atoms tuned to a scattering resonance--- at $T \gtrsim T_F$ provides a rare
platform in which exact scale invariance holds over a
measurable window and multiple independent transport channels
are experimentally accessible.
At unitarity, the interparticle scattering length diverges
and $T$ is the only energy scale, enforcing scale-invariant fluid dynamics.
This allows a non-trivial test of the theorem's central
prediction: that a single $d_{\mathcal{U}}$ governs two physically independent observables---shear viscosity and thermal conductivity---of the same many-body state.

In a scale-invariant fluid, conformal symmetry fixes the frequency dependence of stress fluctuations
---and through the theorem, this frequency dependence is controlled by $d_{\mathcal{U}}$--- so shear viscosity, which measures resistance to flow, obeys $\eta(T) \propto T^{2d_{\mathcal{U}}-2}$.
Cao~\emph{et al.}~\cite{Cao2011} measure $\eta \propto T^{3/2}$
via anisotropic expansion over $E = 2.3$--$4.6\,E_F$, firmly
within the conformal window, giving
\begin{equation}
    2d_{\mathcal{U}} - 2 = \tfrac{3}{2}
    \quad\Rightarrow\quad
    d_{\mathcal{U}} = \tfrac{7}{4}.
\end{equation}
A direct power-law fit yields $d_{\mathcal{U}} = 1.72 \pm 0.03$,
within $1\sigma$ of $7/4$.
Wang \emph{et al.}~\cite{Wang2022}, using a uniform box potential
that eliminates trap-inhomogeneity systematics, independently
recover $d_{\mathcal{U}} = 1.71 \pm 0.07$ from shear viscosity
in the high-temperature subset ($T \gtrsim 0.45\,T_F$).
Both measurements probe the same physical quantity---shear viscosity--- and
their agreement constitutes an internal reproducibility check.

The thermal conductivity $\kappa(T)$, which measures heat transport and is experimentally independent of the viscosity measurements, provides a genuinely independent channel.
Wang \emph{et al.}~\cite{Wang2022} measure $\kappa$ in the same
dataset, finding $d_{\mathcal{U}} = 1.60 \pm 0.09$ from the
same high-temperature cut---consistent with $7/4$ at the
$\lesssim 1.5\sigma$ level and consistent with the shear
channel.
The slight downward shift is expected when the fit window extends
below the strict scale-invariant regime.
Sound diffusivity $D_s$ from Patel \emph{et al.}~\cite{Patel2020}, a composite quantity sensitive to both viscosity and heat transport, provides an additional cross-check.
Restricting to $T \gtrsim 0.75\,T_F$ gives
$d_{\mathcal{U}} = 1.63 \pm 0.13$.

\begin{figure*}[t]
    \centering
    \includegraphics[width=\linewidth]{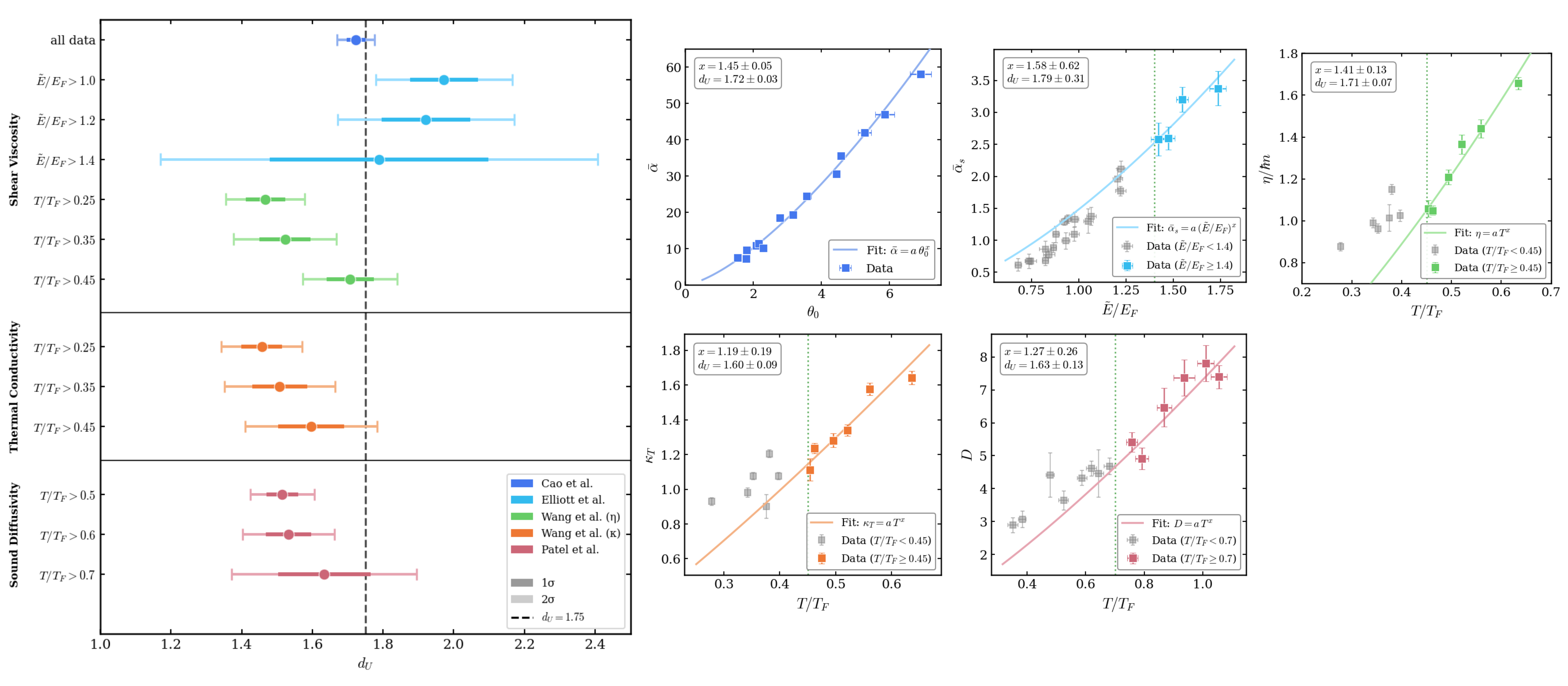}
    \caption{%
    \textbf{Bath-side experimental validation.}
    Consistency check in the unitary Fermi gas. Extracted values of 
    $d_{\mathcal{U}}$ from shear viscosity
    (Cao \emph{et al.}~\protect\cite{Cao2011},
    Elliott \emph{et al.}~\protect\cite{Elliott2014},
    Wang \emph{et al.}~\protect\cite{Wang2022}),
    thermal conductivity
    (Wang \emph{et al.}~\protect\cite{Wang2022}),
    and sound diffusivity
    (Patel \emph{et al.}~\protect\cite{Patel2020}),
    shown under progressively conservative high-$T$ cuts.
    Shear viscosity and thermal conductivity probe independent 
    Green's functions ($G^R_{T_{xy}}$ and $G^R_{J_E J_E}$).
    The dashed line marks $d_{\mathcal{U}} = 7/4$.
    }
    \label{fig:bath_validation}
\end{figure*}

\iffalse
Taken together, two genuinely independent Green's functions
($G^R_{T_{xy}}$ and $G^R_{J_E J_E}$) yield consistent values
of $d_{\mathcal{U}} \approx 7/4$ (Fig.~\ref{fig:bath_validation},
left), validating the theorem's central claim at the level of
spectral scaling.
\fi
Taken together, shear viscosity and thermal conductivity yield mutually consistent, independently obtained values of $d_{\mathcal{U}} \approx 7/4$
(Fig.~\ref{fig:bath_validation}).
We emphasize that neither measurement accesses the spectral
density directly: $d_{\mathcal{U}}$ is inferred from the
temperature scaling of transport coefficients,
under the assumption that this scaling reflects the underlying
power-law structure of environmental fluctuations.
Furthermore, the unitary Fermi gas is conformal only
approximately, in the regime $T \gtrsim T_F$ where $T$ is
the sole energy scale and quantum degeneracy effects are
suppressed; all fits are therefore restricted to this window.
The consistency between independent channels is strong evidence that a single scaling dimension governs the system within this regime,
though we stress that this constitutes a test of the framework under approximate rather than exact scale invariance 

\begin{figure}[t]
    \centering
    \includegraphics[width=\columnwidth]{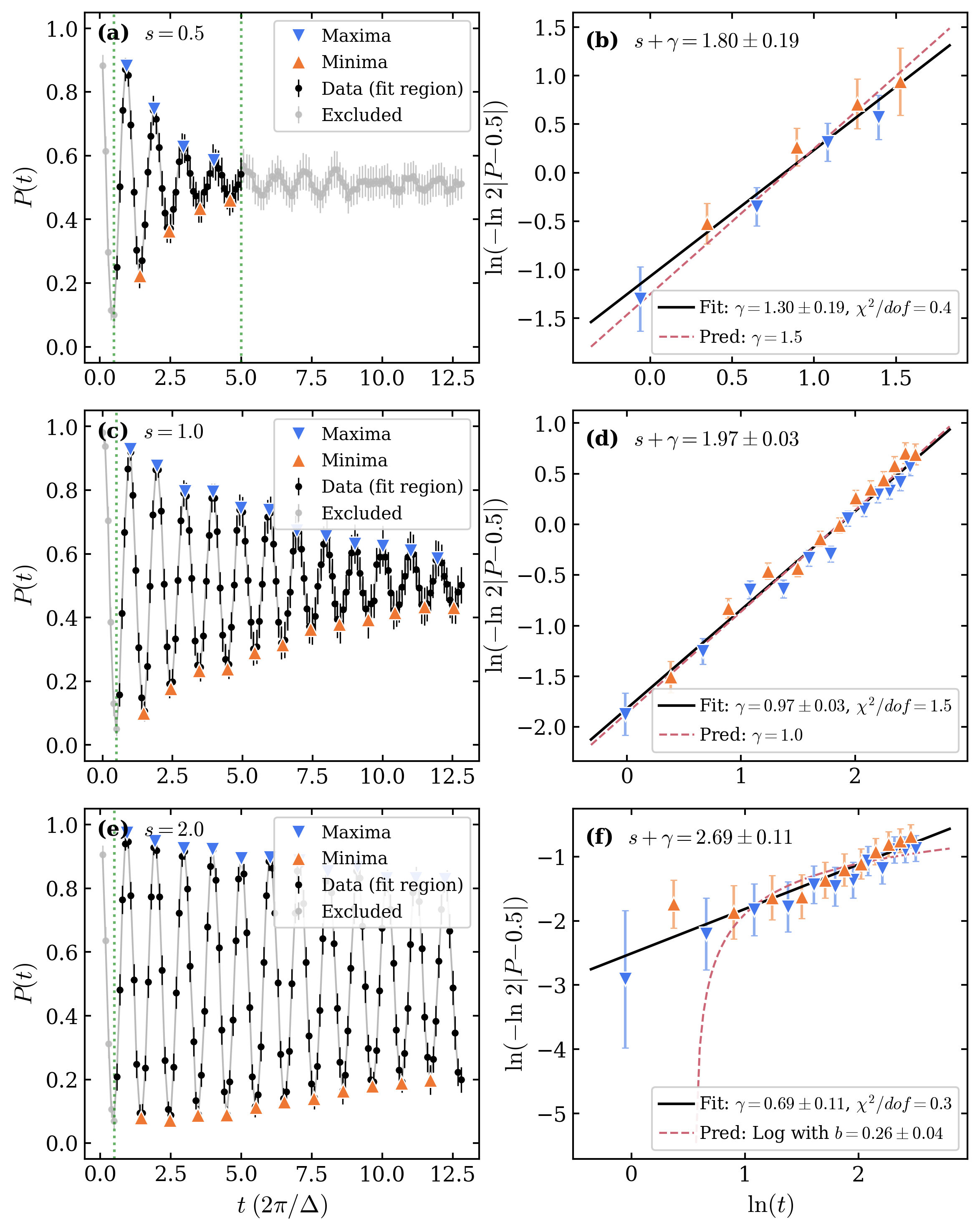}
    \caption{%
    \textbf{Open-system-side validation.}
    Engineered spin-boson baths
    (Sun \emph{et al.}~\protect\cite{Sun:2024obv}).
    Coherence decay envelopes for spectral exponents
    $s = 0.5, 1.0, 2.0$ and corresponding linear fits extracting $\gamma$,
    testing $s + \gamma = 2$.
    Agreement is good for $s = 0.5$ ($1.80\pm0.19$) and
    $s = 1.0$ ($1.97\pm0.03$).
    The $s = 2$ case corresponds to the marginal dimension
    $d_{\mathcal{U}} = 5/2$, where finite bandwidth induces a crossover
    before the asymptotic regime.
    }
    \label{fig:open_validation}
\end{figure}

The open-system side of the theorem is tested independently
by Sun~\emph{et al.}~\cite{Sun:2024obv}, who engineer 
baths with power-law spectral densities $J(\omega) \propto
\omega^s$ in a trapped-ion platform and directly measure
the decoherence of a coupled spin (Fig.~\ref{fig:open_validation},
right).
This is a direct experimental probe of the consistency
relation $s + \gamma = 2$ (Eq.~\eqref{eq:rel1}), independent
of any assumption about the microscopic origin of the bath.
For $s = 0.5$ and $s = 1.0$ we extract $s + \gamma =
1.80 \pm 0.19$ and $1.97 \pm 0.03$, in good agreement with
the prediction.
The $s = 2$ case ($d_{\mathcal{U}} = 5/2$) is the marginal
dimension where asymptotic power-law scaling gives way to
logarithmic growth; finite bandwidth of the engineered bath
accelerates this crossover, and the extracted effective
exponent should be interpreted as pre-asymptotic rather than
a violation of the scaling relation.
We note that these baths are not themselves scale-invariant
--- $J(\omega) \propto \omega^s$ holds only over a narrow
bandwidth --- so this test probes the mathematical structure
of the framework under approximate conditions.
Together with the Fermi gas results---which test the theorem through he properties of the bath itself---, the spin-boson results test it through the response of the coupled system, 
constituting the first two-sided experimental test of the
universality theorem.

Beyond exponent matching, a genuine CFT bath is subject to an
exact constraint from comformal symmetry: the stress tensor must be tracelessness
($T^\mu_\mu = 0$), which implies vanishing bulk viscosity
$\zeta = 0$~\cite{Son2005}---a prediction that no phenomenological power-law bath carries.
A bath engineered to reproduce
$J(\omega) \propto \omega^{1/2}$ matches the power-law behaviour but is free to have any $\zeta$.
Elliott \emph{et al.}~\cite{Elliott2014} measure
$\zeta = 0.005(16)\,\hbar n$, consistent with \emph{exactly}
zero.
This measurement directly tests whether the unitary Fermi gas
is a genuine CFT bath rather than a power-law mimic---and
confirms that it is.

%--------------------------------------------------------------
\section*{Three Physical Realizations}
%--------------------------------------------------------------

The unparticle dimension $d_{\mathcal{U}}$ can be derived from
first principles for any conformal field theory (CFT).
The procedure is systematic: identify the CFT describing the environment, determine the scaling dimension $\Delta$ of the
coupled operator, apply Eq.~\eqref{eq:dU}, and read off all
exponents from Table~\ref{tab:exponents}.
Three realizations spanning 25 orders of magnitude
in energy illustrate the universality.

\paragraph{Quantum Ising criticality (millikelvin scales).}
The two-dimensional quantum Ising CFT has  two relevant primary operators that couple to external probes: the spin
$\sigma$ ($\Delta = 1/8$) and the energy density $\varepsilon$
($\Delta = 1$).
For a probe spin coupled to the energy operator of the
$(2+1)$-dimensional Ising model ($d = 2$):
\begin{equation}
    d_{\mathcal{U}} = 1 - \tfrac{2-2}{2} = 1,
    \quad J(\omega) \propto \omega^{-1}.
\end{equation}
This is the $1/f$ noise spectrum, providing a field-theoretic
explanation for its prevalence near two-dimensional quantum
critical points~\cite{Li:2024zan}.
For the $(1+1)$-dimensional chain coupled to $\varepsilon$:
$d_{\mathcal{U}} = 3/2$, predicting quadratic decoherence growth
$\Gamma_{\mathrm{decoh}} \propto t^2$, testable in trapped-ion
quantum simulators via Ramsey interferometry.

\paragraph{Inflationary cosmology ($10^{13}$ GeV scales).}
In the exponentially expanding spacetime of inflation—well approximated by a $3+1$ de Sitter geometry—a massless scalar field has scaling dimension $\Delta = 3/2$ with respect to de Sitter isometries, which differ from the flat-space conformal group, so the relation~\eqref{eq:dU} must be derived anew in this geometry; the result is
\begin{equation}
    d_{\mathcal{U}} = 2, \quad J(\omega) \propto \omega.
\end{equation}
This is exactly Ohmic---meaning the noise spectrum is linear in frequency--- predicting linear decoherence growth
$\Gamma_{\mathrm{decoh}} \propto Ht$, in agreement with
established results on the quantum-to-classical transition
during inflation~\cite{Kiefer:1998,Colas:2024ysu}.
De Sitter isometries enforce scale invariance even in the
thermal Bunch-Davies vacuum, making inflation the most
robust application of the framework.
Deviations from $d_{\mathcal{U}} = 2$ signal massive fields
or exotic couplings and are constrained by CMB measurements
of the primordial spectral index.

\paragraph{High-energy astrophysical neutrinos (TeV--PeV scales).}
\iffalse
IceCube neutrinos with $E \sim 10^{12}$--$10^{15}$ eV satisfy
$E/T_{\mathrm{CNB}} \sim 10^{16}$--$10^{19}$, placing them
firmly in the vacuum (scale-invariant) regime of any
ambient bath.
If such neutrinos couple to a scale-invariant sector---quantum
gravity, a scale-invariant new sector, or other beyond-Standard-Model
physics---the decoherence rate per unit distance is
\fi
If high-energy astrophysical neutrinos detected by IceCube couple to a scale-invariant sector---quantum
gravity, a conformal fixed point, or other beyond-Standard-Model
physics---the decoherence rate per unit distance is
\begin{equation}
    \frac{\Gamma_{\mathrm{decoh}}}{L}
    \sim \frac{g^2}{M_*^2}\,E^{2d_{\mathcal{U}}-3},
\end{equation}
where $M_*$ is the energy scale suppressing the interactions---for instance the Planck scale for quantum gravity scenarios--- and $g$ is a dimensionless coupling.
The energy dependence is a direct imprint of $d_{\mathcal{U}}$,
making astrophysical neutrino data a powerful probe of scale-invariant new physics.
%A dedicated analysis of IceCube data and future experiments (IceCube-Gen2, KM3NeT) within this framework is in preparation.

IceCube neutrinos with $E \sim 10^{12}$--$10^{15}$~eV have
energies far exceeding the temperature of any scale-invariant sector
thermalized with the diffuse cosmological or intergalactic
environment ($T_{\mathcal{U}} \lesssim \mathrm{keV}$).
The relevant comparison is $E$ versus $T_{\mathcal{U}}$,
the effective temperature of the scale-invariant bath itself,
not that of Standard Model backgrounds such as the CMB or
the cosmic neutrino background.
In this regime, thermal occupation numbers of bath modes at
$\omega \sim E$ are exponentially suppressed, and the
decoherence rate is dominated by its
vacuum contribution, providing a clean imprint of
$d_{\mathcal{U}}$ free of thermal contamination.
For scale-invariant sectors localized near energetic astrophysical
sources such as AGN or GRBs, $T_{\mathcal{U}}$ is a priori
unspecified and the vacuum approximation must be assessed
case by case.
%--------------------------------------------------------------
\section*{Decoherence Phase Transition}
%--------------------------------------------------------------

A striking consequence of the framework is a
\emph{decoherence phase transition} at $d_{\mathcal{U}} = 5/2$.
For $d_{\mathcal{U}} < 5/2$, the decoherence exponent
$\gamma = 5 - 2d_{\mathcal{U}} > 0$ and coherence is
irreversibly lost.
At $d_{\mathcal{U}} = 5/2$, $\gamma = 0$ and decoherence grows
only logarithmically.
For $d_{\mathcal{U}} > 5/2$, $\gamma < 0$:
the decoherence functional \emph{decreases} with time, and
quantum coherence is protected at long times.

This phenomenon is impossible in any memoryless (Markovian) description:
no standard master equation can produce a decoherence functional that saturates or reverses.
Its physical origin is the extremely short correlation time
of super-Ohmic baths---high-frequency modes dominate but their
rapid oscillations average to zero on the system's timescale,
causing the bath to effectively decouple at late times.
The transition is equivalent to the result of
Ref.~\cite{Leggett1987} that coherence survives for $s > 2$
in the spin-boson model, here derived from first principles
as a consequence of conformal symmetry rather than a model-specific calculation.

%--------------------------------------------------------------
\section*{Discussion}
%--------------------------------------------------------------

The unparticle framework does not merely provide a
convenient parametrization of scale-invariant decoherence:
it is its unique characterization.
This elevates the role of $d_{\mathcal{U}}$ to that of an
order parameter for a universality class of non-equilibrium
open quantum systems---analogous to how the central charge $c$
classifies conformal field theories at equilibrium.

The consistency relations~\eqref{eq:rel1}--\eqref{eq:rel3}
are the primary experimental handles.
An experiment that measures, for instance, the noise spectrum
and the decoherence rate independently extracts two values of
$d_{\mathcal{U}}$ and tests their consistency.
Inconsistency is not merely a failure of the model but a
diagnostic: it signals specific physics beyond scale
invariance---a UV cutoff, a mass gap, non-locality, or
multiple competing sectors---and points toward what new
physics is present.

The full fluctuation-dissipation structure that distinguishes
a genuine CFT bath from a power-law mimic requires the
simultaneous absence of a UV cutoff function across the
scaling window.
This is testable in the unitary Fermi gas by verifying that
the thermal equilibrium relation $\tilde\nu(\omega)/\tilde\eta(\omega) =
\coth(\hbar\omega/2k_BT)$---known as the KMS relation--- holds with no measurable rolloff
over the frequency range $\omega \sim k_BT/\hbar$, where
the quantum structure of the $\coth$ factor is operative.
We encourage experimentalists to pursue this measurement as
the definitive test of CFT bath identity.

Three interconnected directions call for further development.
First, the connection to holography: AdS/CFT maps strongly
coupled CFTs to gravitational theories, and the unparticle
bath description of near-horizon fluctuations may provide a
novel link between quantum information and black hole
physics~\cite{Maldacena1999}.
Second, strange metals and spin liquids near quantum critical
points~\cite{Sachdev2011} are natural condensed matter
realizations where the framework's predictions have not yet
been systematically confronted with data.
Third, the extension to gravitational open quantum systems,
where the environment itself may exhibit approximate scale
invariance through asymptotic safety, remains largely
unexplored.

Across all these settings, the unparticle language is not
exotic speculation but the natural effective description
dictated by symmetry---as inevitable, given its assumptions,
as thermodynamics is for equilibrium systems.

%--------------------------------------------------------------
\section*{Methods}
%--------------------------------------------------------------

%--------------------------------------------------------------
\paragraph{Proof of Theorem~\ref{thm:main}.}
%--------------------------------------------------------------

We prove the four conclusions in sequence.

\noindent\textbf{Step 1: Scale invariance implies conformal invariance.}
In a relativistic quantum field theory, the energy-momentum tensor $T^{\mu\nu}$
satisfies $\partial_\mu T^{\mu\nu} = 0$ from translation invariance.
Scale invariance under $x^\mu \to \lambda x^\mu$ requires tracelessness:
$T^\mu_{\ \mu} = 0$.
These two conditions, together with Poincar\'{e} invariance, imply invariance
under the full conformal group $SO(d+1,1)$~\cite{Polchinski1988}, which includes
translations, rotations, dilatations, and special conformal transformations.
The argument fails for two classes of exception, which we exclude by assumption:
theories with exactly marginal operators (beta function identically zero, not just
approximately), and theories with only discrete scale invariance.
For theories with continuous scale invariance, full conformal invariance follows.

\medskip
\noindent\textbf{Step 2: CFT two-point functions are fixed by Ward identities.}
Let $\mathcal{O}_E$ be a primary operator of the CFT with scaling dimension $\Delta$.
Under $x \to \lambda x$ it transforms as
$\mathcal{O}_E(\lambda x) = \lambda^{-\Delta}\mathcal{O}_E(x)$.
Conformal Ward identities fix the Euclidean two-point function up to a single
normalization constant~\cite{DiFrancesco1997}:
\begin{equation}
    \langle \mathcal{O}_E(x)\,\mathcal{O}_E(0)\rangle_{\mathrm{E}}
    = \frac{C_{\mathcal{O}}}{(x^2_E)^\Delta},
    \label{eq:2pt_euclidean}
\end{equation}
where $x^2_E = \sum_\mu (x^\mu)^2$.
This is the unique functional form consistent with conformal symmetry;
no alternative is possible for a primary operator.

\medskip
\noindent\textbf{Step 3: Spectral density from analytic continuation.}
Continuing to Lorentzian signature via $\tau = it - \epsilon$ (Feynman prescription):
\begin{equation}
    \langle \mathcal{O}_E(x,t)\,\mathcal{O}_E(0,0)\rangle_{\mathrm{L}}
    = \frac{C_{\mathcal{O}}}{(-t^2 + \mathbf{x}^2 + i\epsilon)^\Delta}.
\end{equation}
The retarded Green's function at zero spatial momentum ($\mathbf{k} = 0$),
appropriate for a local coupling, is
\begin{equation}
    G_R(\omega,\mathbf{k}=0) \propto (-i\omega)^{2\Delta - d}.
\end{equation}
Dimensional analysis fixes this form; the only freedom is an overall constant.
The spectral density $J(\omega) = -2\,\mathrm{Im}[G_R(\omega)]$ then gives
\begin{equation}
    J(\omega) \propto \omega^{2\Delta - d - 1}, \qquad \omega > 0,
    \label{eq:spectral_step3}
\end{equation}
which is Eq.~\eqref{eq:spectral}.

\medskip
\noindent\textbf{Step 4: Identification with the unparticle form.}
The unparticle spectral density~\cite{Georgi:2007ek} is
$\rho_{\mathcal{U}}(\omega) \propto \omega^{2d_{\mathcal{U}}-3}$.
Matching exponents with Eq.~\eqref{eq:spectral_step3}:
\begin{equation}
    2d_{\mathcal{U}} - 3 = 2\Delta - d - 1
    \quad\Rightarrow\quad
    d_{\mathcal{U}} = \Delta - \tfrac{d-2}{2},
\end{equation}
which is Eq.~\eqref{eq:dU}.
The dynamical exponents in Table~\ref{tab:exponents} follow by Fourier
transformation of the memory kernels
%; full derivations are given in the
%companion paper~\cite{companion}.

%--------------------------------------------------------------
\paragraph{Memory kernels and scaling exponents.}
%--------------------------------------------------------------

Tracing out the environmental degrees of freedom in the path integral yields
a non-Markovian master equation whose dissipation and noise kernels are the
imaginary and real parts of the retarded Green's function, respectively.
Substituting $J(\omega) \propto \omega^{2\Delta - d - 1}$ and applying the
Fourier identity
\begin{equation}
    \int_0^\infty d\omega\; \omega^\mu \sin(\omega t)
    = \Gamma(\mu+1)\sin\!\Bigl[\tfrac{\pi(\mu+1)}{2}\Bigr]\; t^{-(\mu+1)},
\end{equation}
gives the dissipation kernel $\eta(t) \propto t^{-(2d_{\mathcal{U}}-2)}$
and, in the high-temperature limit ($t \gg \beta \equiv 1/T$), the noise kernel
$\nu(t) \propto T\,t^{-(2d_{\mathcal{U}}-3)}$.
Successive time integration yields the damping and decoherence functionals:
\begin{align}
    \Gamma_{\mathrm{damp}}(t)  &\propto t^{3-2d_{\mathcal{U}}}, \\
    \Gamma_{\mathrm{decoh}}(t) &\propto t^{5-2d_{\mathcal{U}}},
\end{align}
establishing all entries of Table~\ref{tab:exponents}.
The consistency relations~\eqref{eq:rel1}--\eqref{eq:rel3} are algebraic
identities among these exponents and require no further proof.
The exact noise kernel valid for all $t$ and $T$ via Matsubara summation.

\paragraph{Loopholes.}
Five classes of physical deviation from the theorem's assumptions are identified.
\emph{(i) Approximate scale invariance}: infrared cutoffs (finite temperature $T$,
system size $L$, mass gap $m$) and ultraviolet cutoffs (lattice spacing, Planck length)
restrict the unparticle description to the scaling window $\omega_{\mathrm{IR}} <
\omega < \omega_{\mathrm{UV}}$; outside this window the spectral density acquires
cutoff functions $f_{\mathrm{IR}}(\omega/\omega_{\mathrm{IR}})$ and
$f_{\mathrm{UV}}(\omega/\omega_{\mathrm{UV}})$ approaching unity in the scaling regime.
\emph{(ii) Non-local coupling}: a coupling kernel $K(x-y)$ introduces a
momentum-dependent form factor modifying the pure power law.
\emph{(iii) Discrete scale invariance}: invariance only under $x \to \lambda_0^n x$
(e.g., Efimov states) produces log-periodic modulations
$J(\omega) \propto \omega^s[1 + A\cos(b\ln\omega + \varphi)]$.
\emph{(iv) Multiple competing sectors}: if the environment contains several unparticle
sectors with dimensions $d_{\mathcal{U}}^{(i)}$, the total spectral density is
$J_{\mathrm{tot}}(\omega) = \sum_i A_i\,\omega^{2d_{\mathcal{U}}^{(i)}-3}$, producing
observable crossovers between sectors.
\emph{(v) Quantum anomalies}: classical scale invariance may be broken by quantum effects
(e.g., the trace anomaly), and the unparticle description applies only above the
anomaly-generated scale.
In all cases, thermal corrections to the noise kernel modify the amplitude but not
the scaling exponents: both the vacuum ($t \ll \beta$) and thermal ($t \gg \beta$)
regimes exhibit power-law behavior with the same $d_{\mathcal{U}}$, differing only
in prefactor.
%Full derivations, including the exact noise kernel via Matsubara summation, are
%given in the companion paper~\cite{companion}.

\paragraph{Contrapositive (no-go theorem).}
The logical contrapositive of the theorem provides an experimental diagnostic:
if the independently measured exponents $s$ and $\gamma_{\mathrm{decoh}}$ are
inconsistent with any single real value of $d_{\mathcal{U}}$
(i.e., $s + \gamma_{\mathrm{decoh}} \neq 2$), then at least one of the following
holds: the environment is not scale-invariant; the coupling is non-local;
Lorentz invariance or unitarity is violated; or multiple competing sectors are
present.
Inconsistency is therefore not a failure of the model but a diagnostic pointing
toward specific new physics.

\paragraph{Stress-tensor channels and extraction of $d_{\mathcal{U}}$.}
In a conformal fluid, the retarded Green's function of the
stress tensor $G^R_{T_{xy}}(\omega) \propto (-i\omega)^{2d_{\mathcal{U}}-1}$
determines the shear viscosity via the Kubo relation
\begin{equation}
    \eta = \lim_{\omega\to 0} \frac{1}{\omega}\,
    \mathrm{Im}\,G^R_{T_{xy}}(\omega),
    \label{eq:kubo_viscosity}
\end{equation}
giving $\eta(T) \propto T^{2d_{\mathcal{U}}-2}$ by dimensional analysis.
Similarly, thermal conductivity is related to the energy-current
correlator $G^R_{J_E J_E}$ via an analogous Kubo relation,
giving $\kappa(T) \propto T^{2d_{\mathcal{U}}-2}$.
The two channels are genuinely independent: they couple to
distinct components of the stress tensor and are measured in
separate experiments.
The unparticle dimension is extracted from power-law fits
$X(T) = a\,(T/T_F)^x$ via
\begin{equation}
    d_{\mathcal{U}} = \frac{x+2}{2},
    \label{eq:dU_fit}
\end{equation}
which follows directly from matching the temperature scaling
to the CFT prediction $\eta(T) \propto T^{2d_{\mathcal{U}}-2}$.

\paragraph{Unitary Fermi gas fits.}
Power-law fits to transport data follow the form
$X(T) = a\,(T/T_F)^x$, with $d_{\mathcal{U}} = (x+2)/2$
for shear viscosity and thermal conductivity.
For Cao \emph{et al.}~\cite{Cao2011} we fit the dimensionless
viscosity coefficient $\bar\alpha$ as a function of the
reduced temperature parameter $\theta_0$, using all data
in the conformal window $E \geq 2.3\,E_F$.
For Wang \emph{et al.}~\cite{Wang2022} and
Patel \emph{et al.}~\cite{Patel2020} we apply conservative
high-$T$ cuts ($T/T_F \geq 0.45$ and $0.75$ respectively)
to restrict to the approximately scale-invariant regime.
Uncertainties are propagated from the original experimental
error bars via orthogonal distance regression.

\paragraph{Decoherence exponent extraction (Sun et al.).}
Coherence decay envelopes are extracted from donor-population
traces by identifying oscillation extrema $A(t) =
|P_{\mathrm{peak}}(t) - 1/2|$ and fitting to a stretched
exponential $A(t) = \tfrac{1}{2}\exp(-Ct^\gamma)$ in the
region $t > 0.5\,(2\pi/\Delta)$, removing the initial
transient.
The exponent $\gamma$ is extracted from the slope of
$\ln(-\ln 2A)$ versus $\ln t$.\footnote{This method relies on the approximation that the exponential envelope varies much more slowly than the oscillations. To check the validity of the approximation, we also perform a full non-linear fit of the data to $P(t) = \frac{1}{2} + \frac{1}{2} e^{-Ct^{\gamma}} \cos(\omega t + \phi)$. With this method, we obtain results that are compatible with those of the linear fit. In the $s=0.5$ case, where the approximation starts being imperfect, the non-linear fit method becomes more precise and yields $\gamma = 1.41 \pm 0.14$, which has a smaller error and is closer to the predicted value of $\gamma=1.5$.}

%--------------------------------------------------------------
\section*{Acknowledgements}
%--------------------------------------------------------------

 GB and HS are supported by PID2023-151418NB-I00 funded by MCIU/AEI/10.13039/501100011033/, and by the European ITN project HIDDeN (H2020-MSCA-ITN-2019/860881-HIDDeN).
GB acknowledges support from the RCCHU. HS is also supported by the grant FPU23/00257, MCIU. TK is supported by the U.S. Department of Energy, Office of Science, Office of Advanced Scientific Computing Research, Department of Energy Computational Science Graduate Fellowship under Award Number DE-SC0025528. The work of GH was supported by the Neutrino Theory Network Fellowship with contract number 726844, and by the U.S. Department of Energy under award number DE-SC0020262.
CA are supported by the Faculty of Arts and Sciences of Harvard University, the National Science Foundation, the Canadian Institute for Advanced Research, the Research Corporation for Science Advancement, the John Templeton Foundation, and the David \& Lucile Packard Foundation. 

%--------------------------------------------------------------
\bibliography{oqs_bib}

\begin{thebibliography}{10}
\expandafter\ifx\csname url\endcsname\relax
  \def\url#1{\texttt{#1}}\fi
\expandafter\ifx\csname urlprefix\endcsname\relax\def\urlprefix{URL }\fi
\providecommand{\bibinfo}[2]{#2}
\providecommand{\eprint}[2][]{\url{#2}}

\bibitem{Georgi:2007ek}
\bibinfo{author}{Georgi, H.}
\newblock \bibinfo{title}{{Unparticle physics}}.
\newblock \emph{\bibinfo{journal}{Phys. Rev. Lett.}} \textbf{\bibinfo{volume}{98}}, \bibinfo{pages}{221601} (\bibinfo{year}{2007}).
\newblock \eprint{hep-ph/0703260}.

\bibitem{Lindblad1976}
\bibinfo{author}{Lindblad, G.}
\newblock \bibinfo{title}{On the generators of quantum dynamical semigroups}.
\newblock \emph{\bibinfo{journal}{Communications in Mathematical Physics}} \textbf{\bibinfo{volume}{48}}, \bibinfo{pages}{119--130} (\bibinfo{year}{1976}).
\newblock \bibinfo{note}{GKSL master equation for open quantum systems}.

\bibitem{CaldeiraLeggett1983}
\bibinfo{author}{Caldeira, A.~O.} \& \bibinfo{author}{Leggett, A.~J.}
\newblock \bibinfo{title}{Quantum tunnelling in a dissipative system}.
\newblock \emph{\bibinfo{journal}{Annals of Physics}} \textbf{\bibinfo{volume}{149}}, \bibinfo{pages}{374--456} (\bibinfo{year}{1983}).

\bibitem{Polchinski1988}
\bibinfo{author}{Polchinski, J.}
\newblock \emph{\bibinfo{title}{String Theory. Vol. 1: An Introduction to the Bosonic String}} (\bibinfo{publisher}{Cambridge University Press}, \bibinfo{address}{Cambridge}, \bibinfo{year}{1988}).

\bibitem{Nakayama2015}
\bibinfo{author}{Nakayama, Y.}
\newblock \bibinfo{title}{Scale invariance vs conformal invariance}.
\newblock \emph{\bibinfo{journal}{Phys. Rept.}} \textbf{\bibinfo{volume}{569}}, \bibinfo{pages}{1--93} (\bibinfo{year}{2015}).
\newblock \eprint{1302.0884}.

\bibitem{Luty2013}
\bibinfo{author}{Luty, M.~A.}, \bibinfo{author}{Polchinski, J.} \& \bibinfo{author}{Rattazzi, R.}
\newblock \bibinfo{title}{The $a$-theorem and the asymptotics of 4d quantum field theory}.
\newblock \emph{\bibinfo{journal}{JHEP}} \textbf{\bibinfo{volume}{01}}, \bibinfo{pages}{152} (\bibinfo{year}{2013}).
\newblock \eprint{1204.5221}.

\bibitem{Cao2011}
\bibinfo{author}{Cao, C.}, \bibinfo{author}{Elliott, E.}, \bibinfo{author}{Wu, J.} \& \bibinfo{author}{Thomas, J.~E.}
\newblock \bibinfo{title}{Universal quantum viscosity in a unitary {F}ermi gas}.
\newblock \emph{\bibinfo{journal}{Science}} \textbf{\bibinfo{volume}{331}}, \bibinfo{pages}{58--61} (\bibinfo{year}{2011}).
\newblock \eprint{1007.2625}.

\bibitem{Wang2022}
\bibinfo{author}{Wang, H.}, \bibinfo{author}{Dong, Y.} \& \bibinfo{author}{Thomas, J.~E.}
\newblock \bibinfo{title}{Universal sound diffusion in a strongly interacting fermi gas}.
\newblock \emph{\bibinfo{journal}{Phys. Rev. Lett.}} \textbf{\bibinfo{volume}{128}}, \bibinfo{pages}{090402} (\bibinfo{year}{2022}).

\bibitem{Patel2020}
\bibinfo{author}{Patel, P.~B.} \emph{et~al.}
\newblock \bibinfo{title}{Universal sound diffusion in a strongly interacting fermi gas}.
\newblock \emph{\bibinfo{journal}{Science}} \textbf{\bibinfo{volume}{370}}, \bibinfo{pages}{1222--1226} (\bibinfo{year}{2020}).
\newblock \urlprefix\url{https://www.science.org/doi/abs/10.1126/science.aaz5756}.

\bibitem{Sun:2024obv}
\bibinfo{author}{Sun, K.} \emph{et~al.}
\newblock \bibinfo{title}{{Quantum simulation of spin-boson models with structured bath}}.
\newblock \emph{\bibinfo{journal}{Nature Commun.}} \textbf{\bibinfo{volume}{16}}, \bibinfo{pages}{4042} (\bibinfo{year}{2025}).
\newblock \eprint{2405.14624}.

\bibitem{Son2005}
\bibinfo{author}{Son, D.~T.}
\newblock \bibinfo{title}{Vanishing bulk viscosities and conformal invariance of the unitary {F}ermi gas}.
\newblock \emph{\bibinfo{journal}{Phys. Rev. Lett.}} \textbf{\bibinfo{volume}{98}}, \bibinfo{pages}{020604} (\bibinfo{year}{2007}).
\newblock \eprint{cond-mat/0511721}.

\bibitem{Elliott2014}
\bibinfo{author}{Elliott, E.}, \bibinfo{author}{Joseph, J.~A.} \& \bibinfo{author}{Thomas, J.~E.}
\newblock \bibinfo{title}{Observation of conformal symmetry breaking and scale invariance in a strongly interacting {F}ermi gas}.
\newblock \emph{\bibinfo{journal}{Phys. Rev. Lett.}} \textbf{\bibinfo{volume}{112}}, \bibinfo{pages}{040405} (\bibinfo{year}{2014}).
\newblock \eprint{1308.3162}.

\bibitem{Li:2024zan}
\bibinfo{author}{Li, Y.} \emph{et~al.}
\newblock \bibinfo{title}{{Critical fluctuations and noise spectra in two-dimensional Fe$_{3}$GeTe$_{2}$ magnets}}.
\newblock \emph{\bibinfo{journal}{Nature Commun.}} \textbf{\bibinfo{volume}{16}}, \bibinfo{pages}{8585} (\bibinfo{year}{2025}).
\newblock \eprint{2407.00647}.

\bibitem{Kiefer:1998}
\bibinfo{author}{Kiefer, C.}, \bibinfo{author}{Polarski, D.} \& \bibinfo{author}{Starobinsky, A.~A.}
\newblock \bibinfo{title}{Quantum-to-classical transition for fluctuations in the early universe}.
\newblock \emph{\bibinfo{journal}{International Journal of Modern Physics D}} \textbf{\bibinfo{volume}{7}}, \bibinfo{pages}{455--462} (\bibinfo{year}{1998}).
\newblock \eprint{gr-qc/9802003}.

\bibitem{Colas:2024ysu}
\bibinfo{author}{Colas, T.}, \bibinfo{author}{Grain, J.}, \bibinfo{author}{Kaplanek, G.} \& \bibinfo{author}{Vennin, V.}
\newblock \bibinfo{title}{{In-in formalism for the entropy of quantum fields in curved spacetimes}}.
\newblock \emph{\bibinfo{journal}{JCAP}} \textbf{\bibinfo{volume}{08}}, \bibinfo{pages}{047} (\bibinfo{year}{2024}).
\newblock \eprint{2406.17856}.

\bibitem{Leggett1987}
\bibinfo{author}{Leggett, A.~J.} \emph{et~al.}
\newblock \bibinfo{title}{Dynamics of the dissipative two-state system}.
\newblock \emph{\bibinfo{journal}{Reviews of Modern Physics}} \textbf{\bibinfo{volume}{59}}, \bibinfo{pages}{1--85} (\bibinfo{year}{1987}).

\bibitem{Maldacena1999}
\bibinfo{author}{Maldacena, J.}
\newblock \bibinfo{title}{The large n limit of superconformal field theories and supergravity}.
\newblock \emph{\bibinfo{journal}{International Journal of Theoretical Physics}} \textbf{\bibinfo{volume}{38}}, \bibinfo{pages}{1113--1133} (\bibinfo{year}{1999}).
\newblock \eprint{hep-th/9711200}.

\bibitem{Sachdev2011}
\bibinfo{author}{Sachdev, S.}
\newblock \emph{\bibinfo{title}{Quantum Phase Transitions}} (\bibinfo{publisher}{Cambridge University Press}, \bibinfo{year}{2011}), \bibinfo{edition}{2} edn.

\bibitem{DiFrancesco1997}
\bibinfo{author}{Di~Francesco, P.}, \bibinfo{author}{Mathieu, P.} \& \bibinfo{author}{S{\'e}n{\'e}chal, D.}
\newblock \emph{\bibinfo{title}{Conformal Field Theory}} (\bibinfo{publisher}{Springer}, \bibinfo{address}{New York}, \bibinfo{year}{1997}).

\end{thebibliography}
%--------------------------------------------------------------

\end{document}